\pdfoutput=1

\documentclass[acmsmall, manuscript, screen, review=False]{acmart}

\usepackage{wrapfig}
\usepackage{subcaption}

\acmConference[CSCW '24]{The 27th ACM SIGCHI Conference on Computer-Supported Cooperative Work \& Social Computing (CSCW)}{November 9-13, 2024}{San Jose, Costa Rica}

\title[Bridging Dictionary]{Bridging Dictionary: AI-Generated Dictionary of Partisan Language Use}

\author{Hang Jiang*}
\email{hjian42@mit.edu}
\orcid{0009-0000-9523-136X}
\affiliation{
  \institution{MIT Center for Constructive Communication \& MIT Media Lab}
  \city{Cambridge}
  \state{MA}
  \country{USA}
}

\author{Doug Beeferman*}
\email{dougb5@mit.edu}
\orcid{0009-0005-5879-5744}
\affiliation{
  \institution{MIT Center for Constructive Communication \& MIT Media Lab}
  \city{Cambridge}
  \state{MA}
  \country{USA}
}

\author{William Brannon}
\email{wbrannon@mit.edu}
\orcid{0000-0002-1435-8535}
\affiliation{
  \institution{MIT Center for Constructive Communication \& MIT Media Lab}
  \city{Cambridge}
  \state{MA}
  \country{USA}
}

\author{Andrew Heyward}
\email{aheyward@media.mit.edu}
\affiliation{
  \institution{MIT Center for Constructive Communication \& MIT Media Lab}
  \city{Cambridge}
  \state{MA}
  \country{USA}
}

\author{Deb Roy}
\email{dkroy@mit.edu}
\orcid{0000-0002-2780-4768}
\affiliation{
  \institution{MIT Center for Constructive Communication \& MIT Media Lab}
  \city{Cambridge}
  \state{MA}
  \country{USA}
}



\begin{document}

\begin{abstract}
Words often carry different meanings for people from diverse backgrounds. Today's era of social polarization demands that we choose words carefully to prevent miscommunication, especially in political communication and journalism. To address this issue, we introduce the Bridging Dictionary, an interactive tool designed to illuminate how words are perceived by people with different political views. The Bridging Dictionary includes a static, printable document featuring 796 terms with summaries generated by a large language model. These summaries highlight how the terms are used distinctively by Republicans and Democrats. Additionally, the Bridging Dictionary offers an interactive interface that lets users explore selected words, visualizing their frequency, sentiment, summaries, and examples across political divides. We present a use case for journalists and emphasize the importance of human agency and trust in further enhancing this tool\footnote{HJ and DB are both co-first authors. HJ led the evaluation and writing and DB led the tool development.}. The deployed version of Bridging Dictionary is available at \url{https://dictionary.ccc-mit.org/}.
\end{abstract}

\maketitle

\section{Introduction}
\label{sec:introduction}
Polarization is a significant feature of the political landscape in the United States \cite{poole1984polarization, mccarty2016polarized, heltzel2020polarization}. Previous research has shown that Republicans and Democrats often use and interpret words differently, even when speaking the same language \cite{li2017speaking, notthesame}. This linguistic divide poses considerable challenges to public discourse, particularly in journalism, where the use of words without an understanding of their varying connotations across political communities can have serious consequences. Journalists face the added difficulty of manually reading and editing news content, a process that is not only time-consuming but also prone to errors due to the constantly evolving connotations of words online. To address this problem, we introduce the Bridging Dictionary (BD), a tool designed to automatically identify controversial terms across the political divides and to summarize their different usages. We provide not only a useful resource for the academic community, journalists, and wider audiences but also highlight the importance of considering human agency and trust in developing human-AI systems.  


\section{Related Work}
\label{sec:related-work}

\paragraph{Polarized language use in NLP} Researchers in political science and Natural Language Processing (NLP) have found that there is a partisan difference in language understanding \cite{li2017speaking}. \citet{notthesame} used modern machine-translation methods to show that the Republican and Democratic communities use English words differently. For instance, there are different connotations when partisans use ``undocumented workers'' or ``illegal aliens'' to discuss the same group of people. To address this issue, \citet{webson-etal-2020-undocumented} developed an NLP method to mitigate the political bias of text representations, showing that it improves the viewpoint diversity of document rankings. Recently, NLP researchers have also proposed novel methods to quantify and debias the political bias of language models \cite{liu2021mitigating,liu2022quantifying}. However, previous work tends to focus on measuring and debiasing NLP models instead of facilitating humans in writing less biased content. Our work fills the gap by leveraging NLP to help humans understand the language bias and facilitate them in writing and editing through an interactive tool. 

\paragraph{NLP for qualitative analysis and sensemaking} NLP has been used to develop computational tools for qualitative analysis and sensemaking \cite{crowston2012using,guetterman2018augmenting}. Among these tools, topic models are particularly prevalent for text analysis across various domains \cite{yan2013biterm,hu2019hotel,isoaho2021topic}. However, traditional NLP models often lack world knowledge, resulting in limited insights and interpretability \cite{barde2017overview}. The modern large language models (LLMs) have enabled users to interact with unstructured data through simple queries to extract more nuanced and interpretable insights. Recent studies have explored the application of LLMs to automatically extract and summarize valuable information from texts \cite{chew2023llm,xiao2023supporting,gero2024supporting,overneySenseMateAccessibleBeginnerFriendly2024}. Despite these advancements, there is a lack of research focusing on how LLMs can assist journalists by providing qualitative insights for writing and editing. This study aims to bridge this gap by introducing an interactive tool to summarize the varying usage of terms across political divides, thereby guiding journalists in their word choices in news writing.

\section{System Overview}
\label{sec:system-overview}

\begin{figure}[ht!]
    \centering
    \begin{subfigure}[b]{0.45\textwidth}
        \centering
        \includegraphics[width=\textwidth]{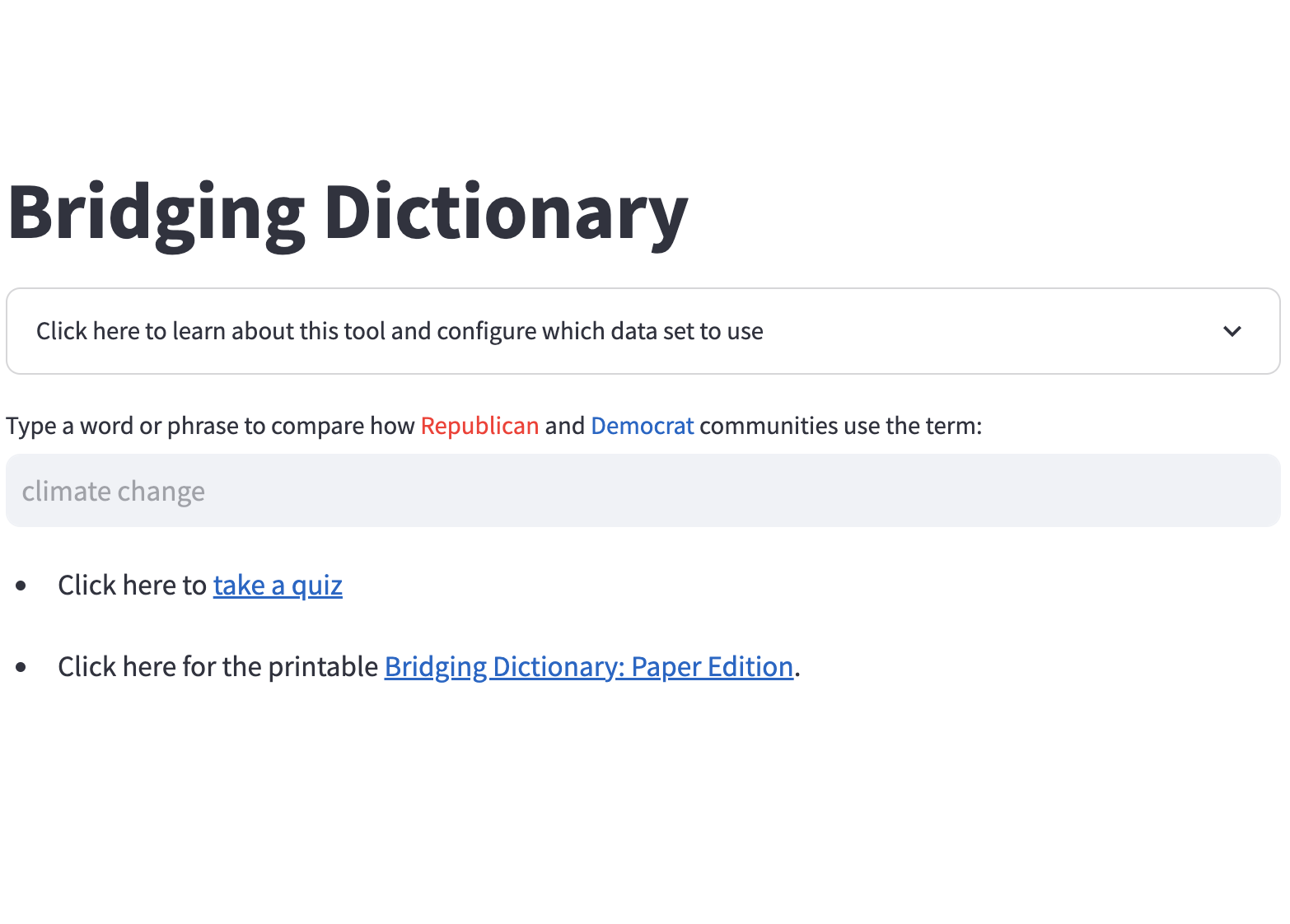}
        \caption{The front page of the interactive Bridging Dictionary demo. The user can type any term they are interested in.}
        \label{fig:bd_page}
    \end{subfigure}
    \hfill 
    \begin{subfigure}[b]{0.50\textwidth}
        \centering
        \includegraphics[width=\textwidth]{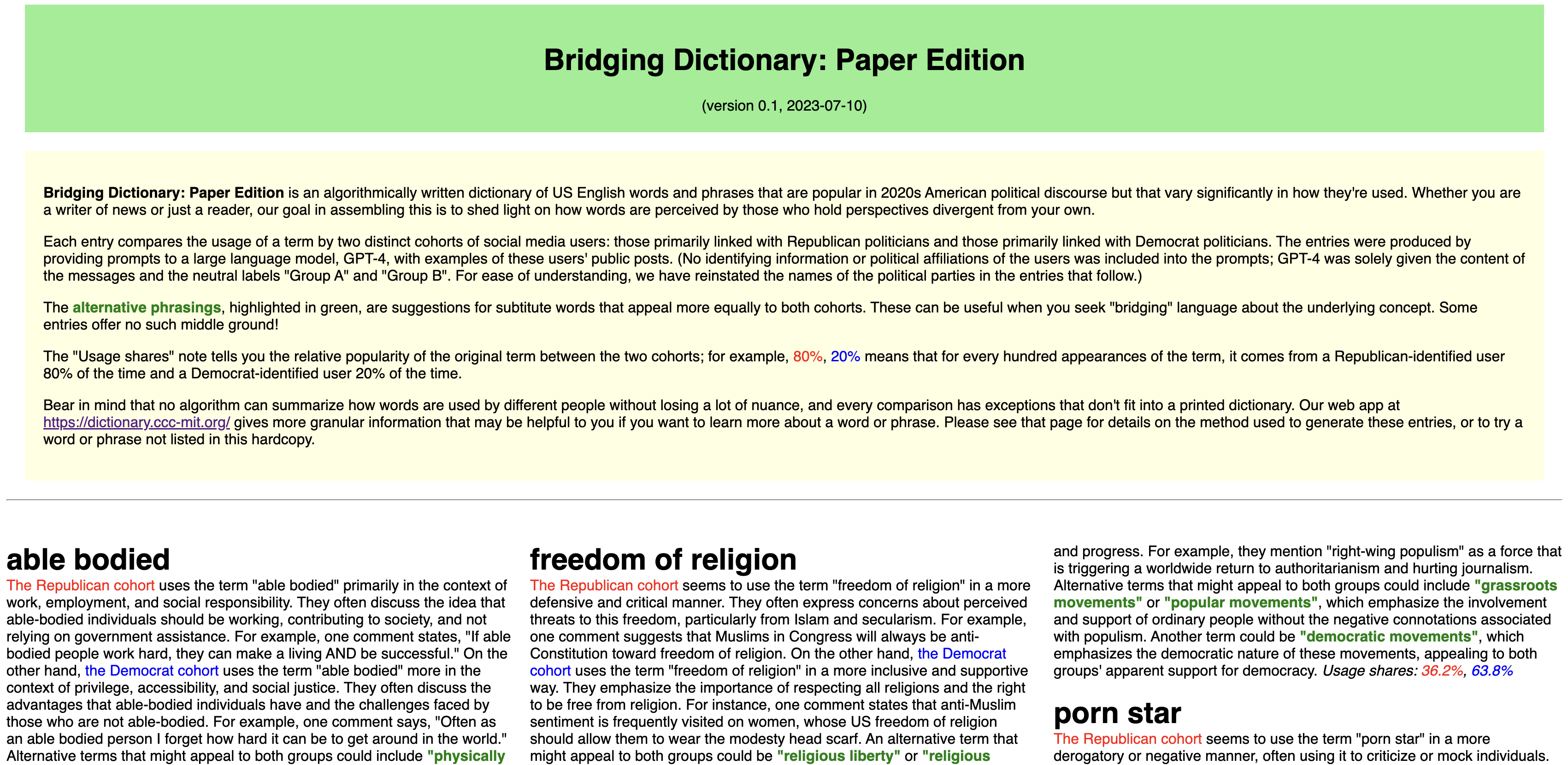}
        \caption{The front page of ``Bridging Dictionary: Paper Edition'' with GPT-generated summaries for representative terms.}
        \label{fig:bd_paper_edition}
    \end{subfigure}
    \caption{Selected views of the Bridging Dictionary: an interactive front page with user text input, and a static paper edition page.}
    \label{fig:interface}
\end{figure}

The Bridging Dictionary (BD) comprises two main components: (1) a paper dictionary and (2) an interactive demo. As illustrated in Figure \ref{fig:interface}, the paper dictionary presents 796 representative terms in a print-ready format, supplemented with summaries generated by an LLM. The interactive demo, on the other hand, enables users to explore the usage of a given term within Republican and Democratic communities in greater detail. BD leverages \texttt{gpt-3.5-turbo}, a widely-recognized LLM, via OpenAI's API to generate these summaries. The term usages are sampled from a Twitter dataset \cite{jiang-etal-2022-communitylm}, which includes 4.7 million partisan-generated tweets (amounting to 100 million tokens) from each side during the 2020 American election. Users can customize both the dataset and the available LLMs from OpenAI.

\subsection{Paper Dictionary}

We generate a static printable document called ``Bridging Dictionary: Paper Edition'' that comprises 796 terms with LLM-generated summaries. These terms are curated algorithmically: we identify words and multi-word phrases that (1) occur sufficiently often within both partisan communities and (2) have significant differences between the two communities, either in sentiment score or in usage frequency. The parameters for these operations (i.e., the thresholds for ``sufficient'' and ``significant'') are adjusted manually by an editor.

\subsection{Interactive Demo}

The interactive demo is a web-based generative dictionary providing greater detail and flexibility than the print version.  It is implemented with the Streamlit framework \cite{streamlitStreamlitFasterWay2021}. Whenever a user types a phrase, the system provides a few functions to explore how two communities (Republicans and Democrats) use this term, with alternative suggestions and visualizations of the input data.

\begin{figure}[htb]
    \centering 
\begin{subfigure}{0.4\textwidth}
  \includegraphics[width=\linewidth]{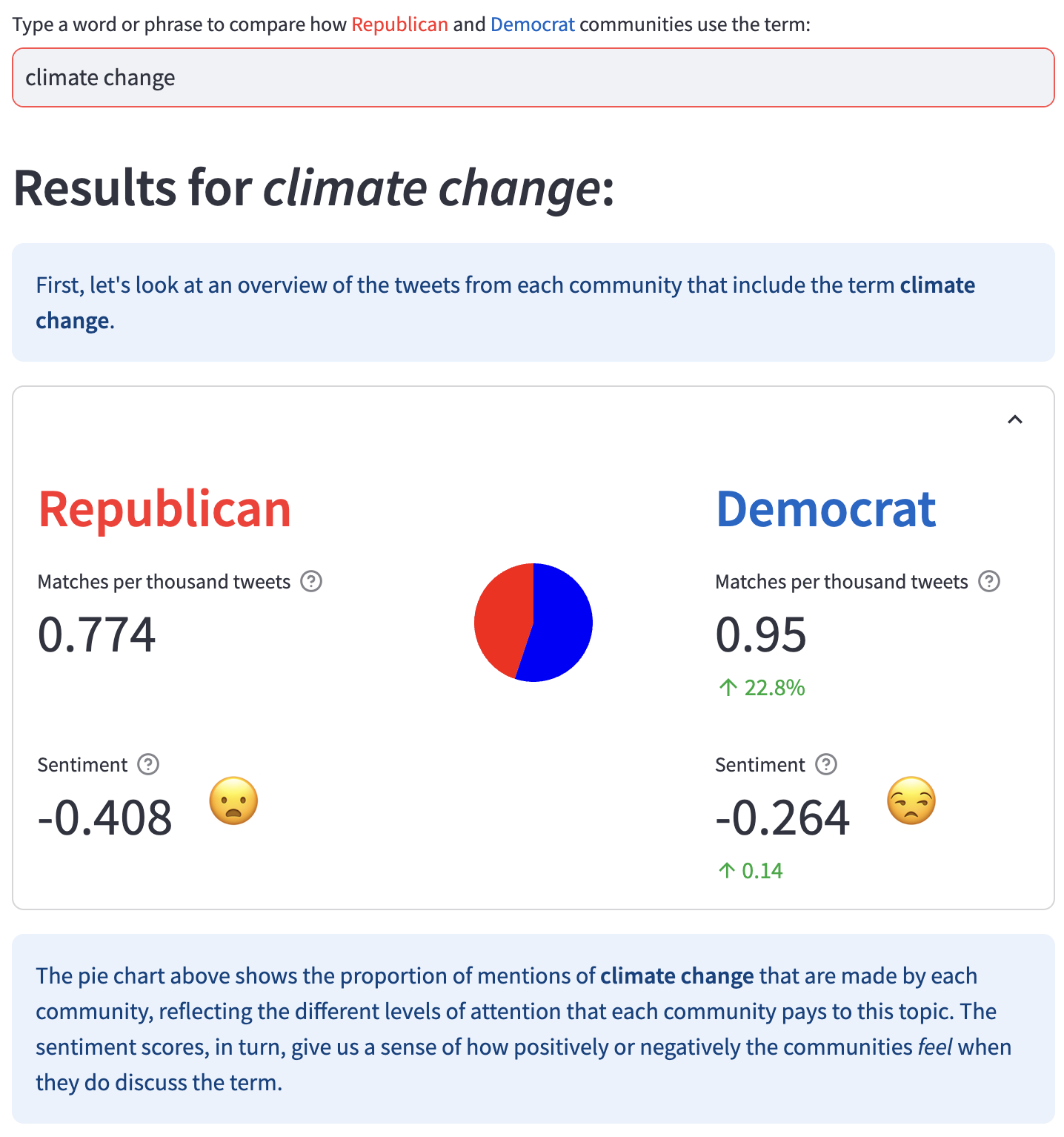}
  \caption{Statistics}
  \label{fig:1}
\end{subfigure}
\begin{subfigure}{0.4\textwidth}
  \includegraphics[width=\linewidth]{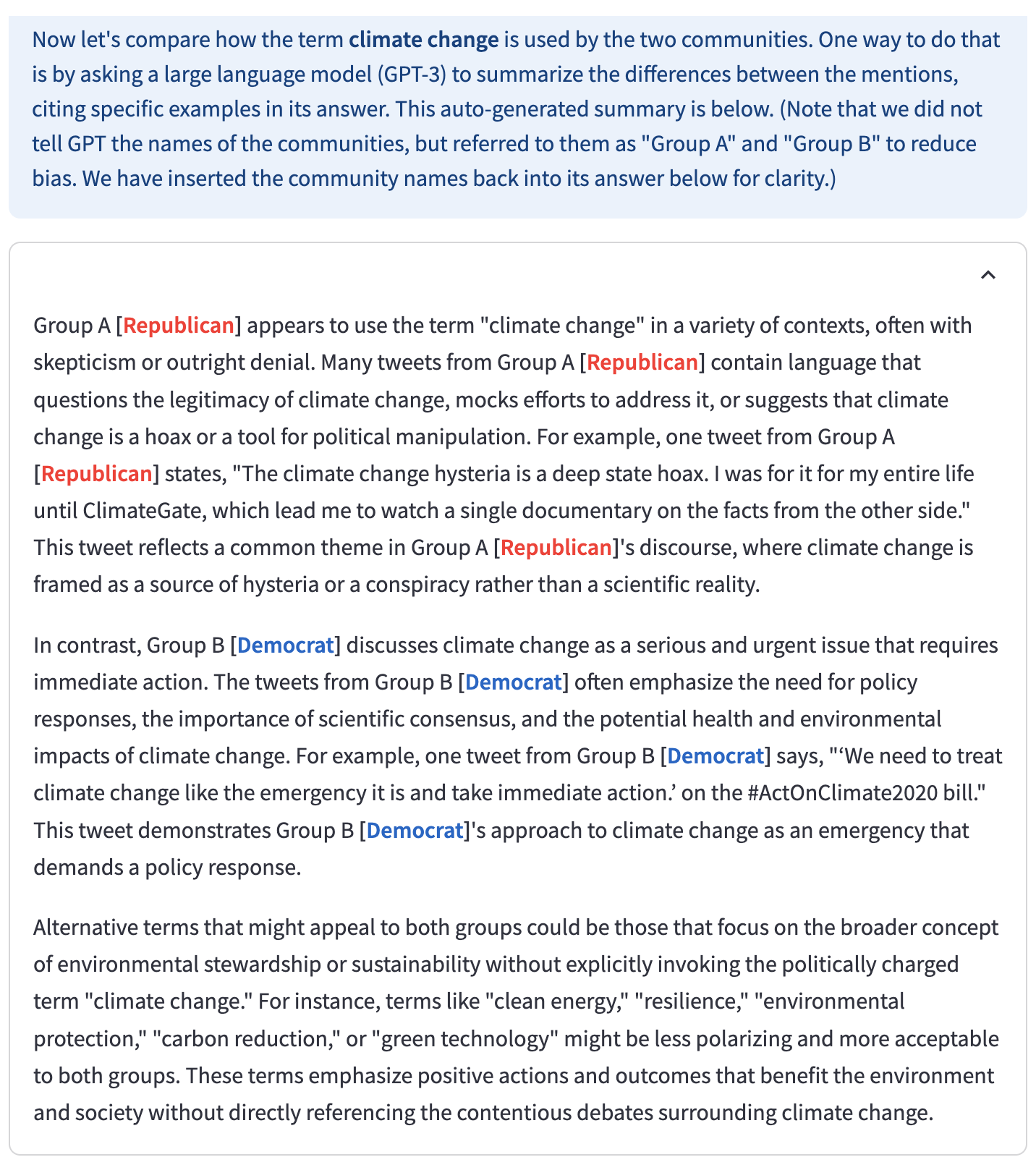}
  \caption{Summary}
  \label{fig:2}
\end{subfigure}

\medskip
\begin{subfigure}{0.33\textwidth}
\includegraphics[width=\linewidth]{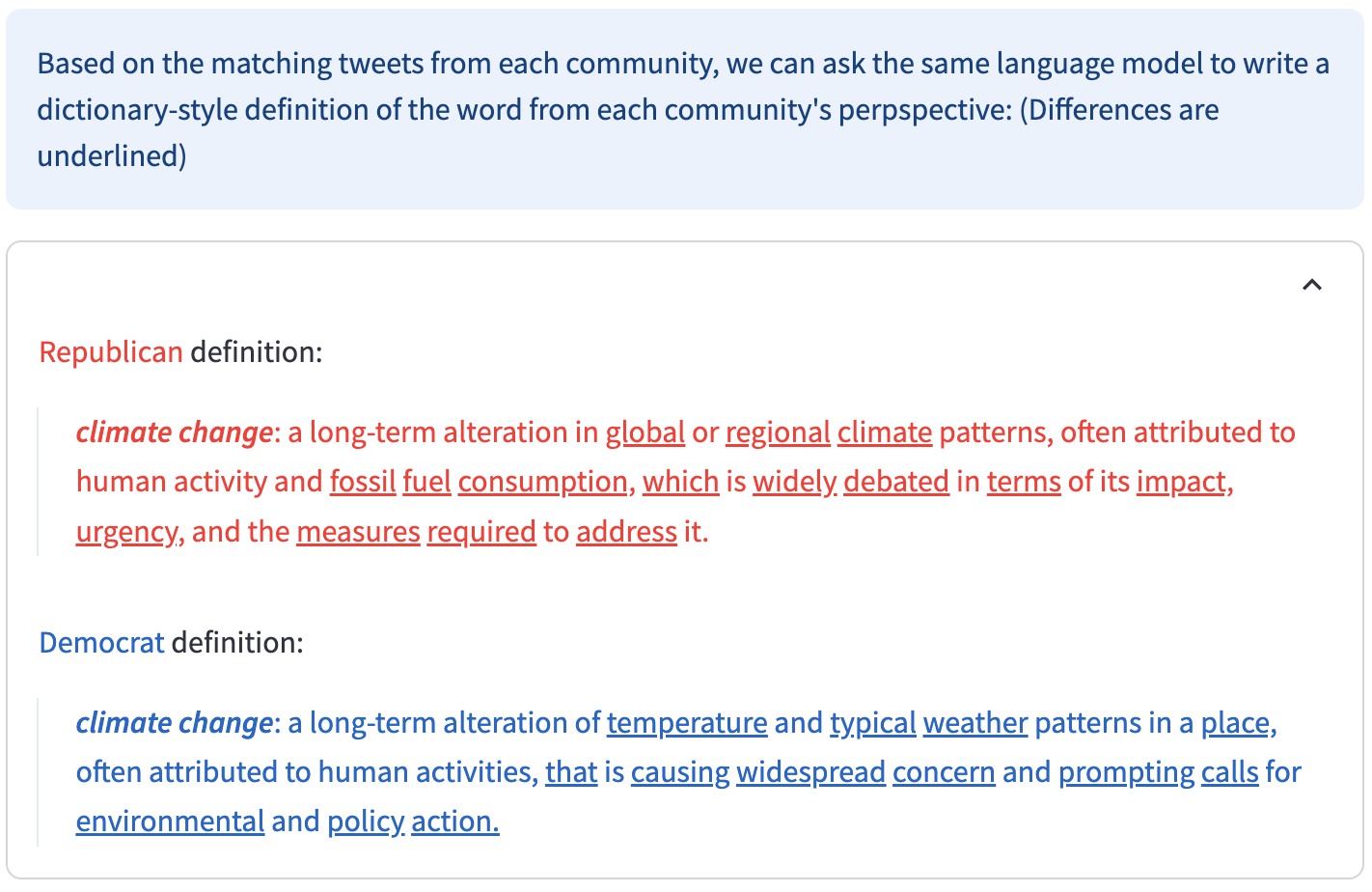}
  \caption{Definition}
  \label{fig:3}
\end{subfigure}\hfil 
\begin{subfigure}{0.33\textwidth}
  \includegraphics[width=\linewidth]{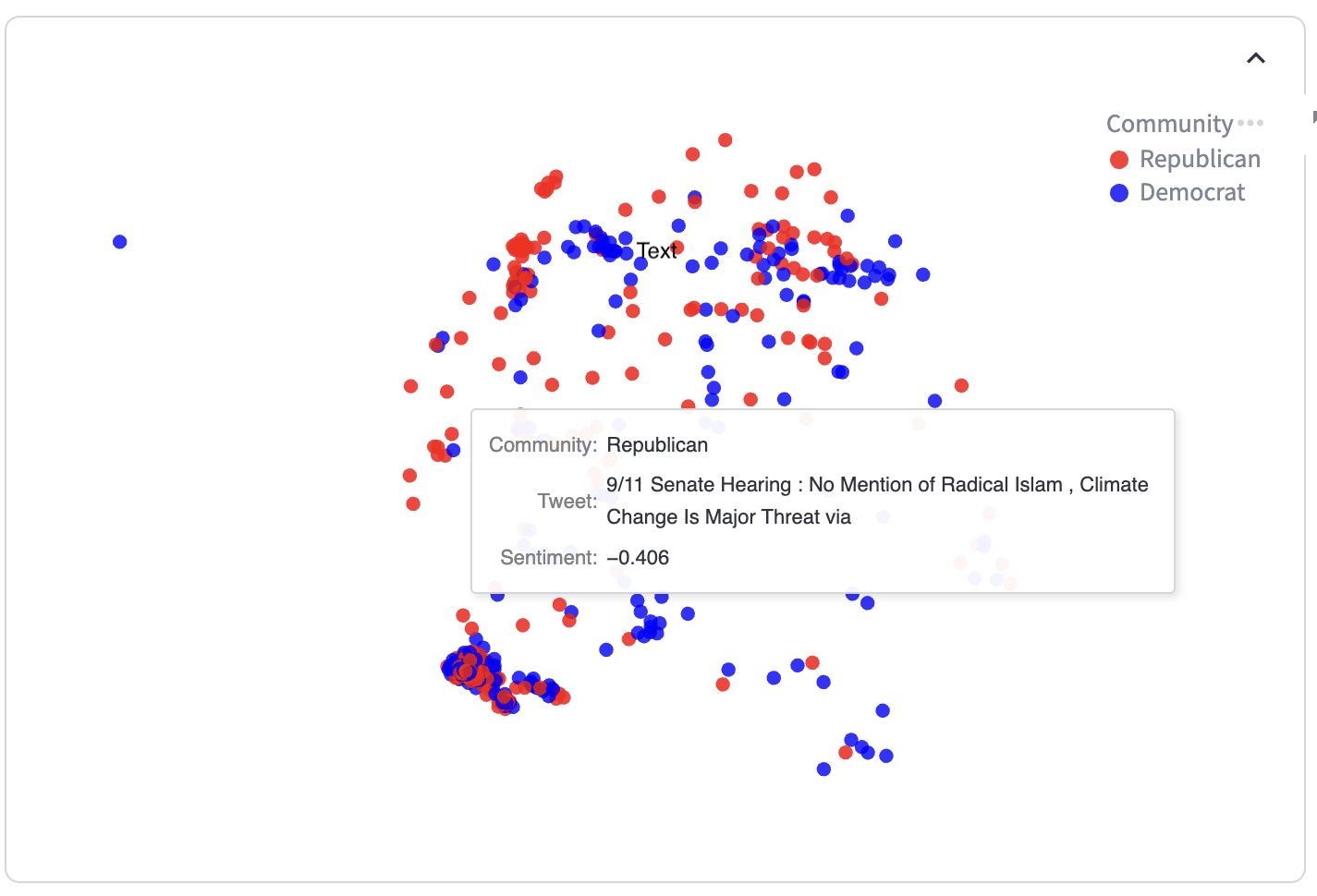}
  \caption{Topic scatterplot}
  \label{fig:4}
\end{subfigure}\hfil 
\begin{subfigure}{0.33\textwidth}
  \includegraphics[width=\linewidth]{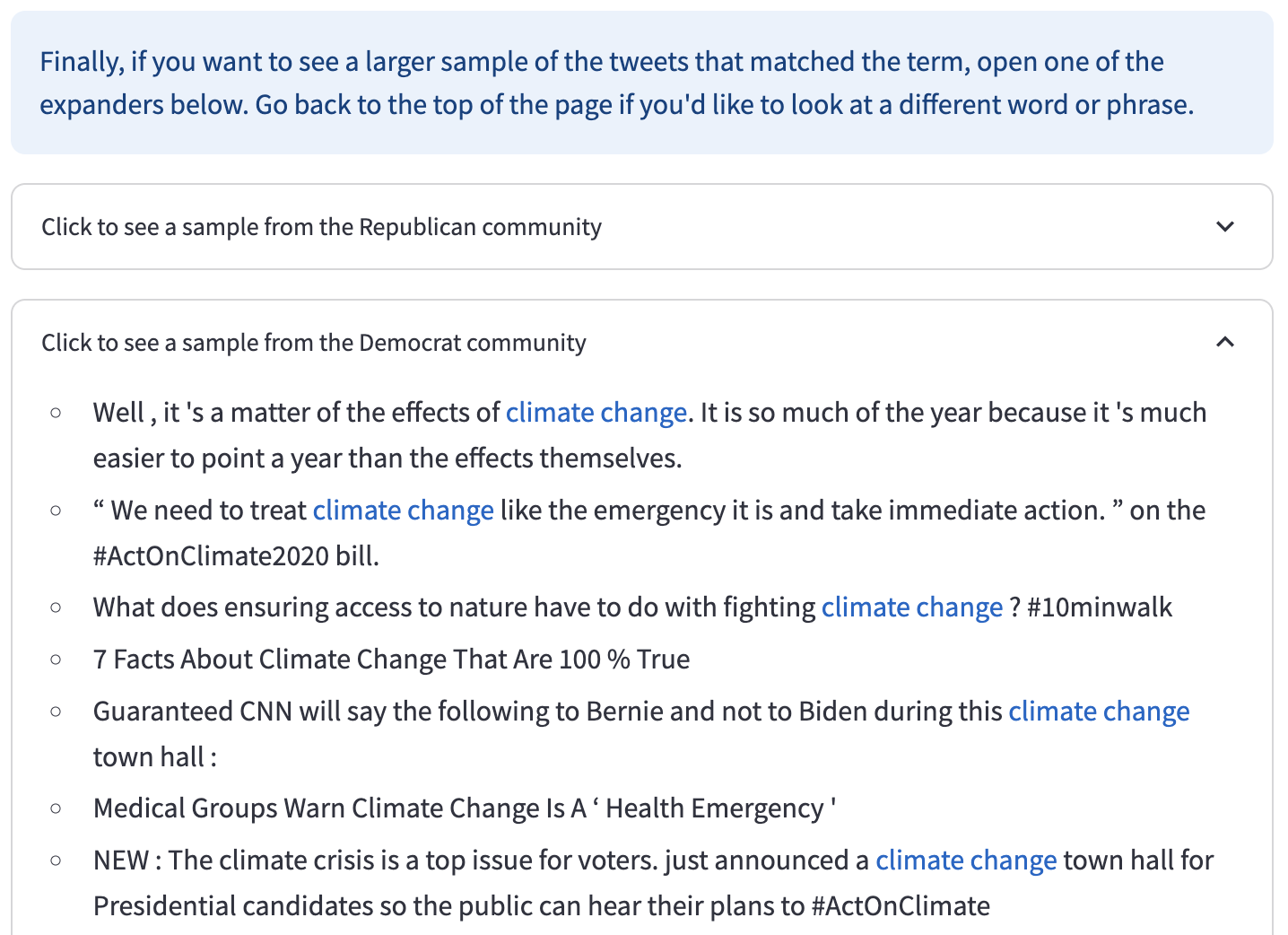}
  \caption{Sample list}
  \label{fig:6}
\end{subfigure}
\caption{Six core sections in the interactive demo to help users explore how two communities use a term differently.}
\label{fig:images}
\end{figure}

\textit{Statistics.} This section offers an overview of tweets containing the specified term from each community. The column ``Matches per Thousand Tweets'' indicates the frequency of tweets that include the term (e.g., ``climate change'') per 1,000 tweets within a community. The accompanying pie chart illustrates the proportion of term usage between two partisan communities. Sentiment scores represent the average sentiment for tweets from a community that matched the term, with higher scores indicating a more positive sentiment. Colored text highlights the comparative scores between the two communities, helping users to interpret the data easily.

\textit{Summary.} This section features LLM-generated summaries that explain the usage of the term across divides and propose alternative terms. The generation follows a standard retrieval-augmented generation (RAG) procedure, as outlined by \citet{Gao2023RetrievalAugmentedGF}. Initially, the system randomly samples up to 50 tweets containing the term from a specific community, creates a prompt using these tweets, and prompts the LLM to produce summaries with a simple query. Importantly, the LLM is unaware of the community identity and only uses the sampled tweets for its summarization.

\textit{Definition.} This section generates a dictionary-style definition of a term from each community's perspective. This follows the same RAG process as the Summary section but asks the model for a definition instead of a summary.

\textit{Topic scatterplot.} This section presents a two-dimensional interactive scatterplot that organizes individual tweets by topic, grouping similar topics closely together. Users can explore specific tweets by hovering over the corresponding points, enabling them to review the sampled tweets used for summary and definition generation. The process follows a well-established pipeline (e.g., BERTopic \cite{grootendorst2022bertopic}) transforming raw text into a scatterplot. It involves computing the embedding for each tweet using a sentence embedding model, then projecting these embeddings into two dimensions. By default, the points are clustered and color-coded based on the outcome of a clustering algorithm applied to these embeddings. For projection, we employ UMAP \cite{mcinnes2018umap}, and for clustering, we use HDBSCAN \cite{mcinnes2017hdbscan}.



\textit{Sample list.} This section lists the sampled tweets from each group, allowing users to read the information source.

\section{Discussion and Evaluation}
\label{sec:discussion}

After deploying the Bridging Dictionary, we interviewed a professional journalist from Frontline at PBS and received positive feedback. Based on the interview, our statistics and summary features significantly aid in understanding the political divide in language use. Additionally, the topic scatterplot and sample list features are relied upon for supporting LLM-generated content and assisting journalists in making informed word choices. During our interview, two promising directions emerged: (1) enhancing the connection between LLM-generated content and information sources by considering human agency and trust in human-AI interaction, and (2) broadening the range of information sources beyond Twitter and regularly updating the dataset to reflect the evolving nature of language across different platforms and over time. We intend to further develop the tool based on these suggestions, conduct a more comprehensive field study involving more professional journalists and other users, and assess the tool's impact on writing and editing.



\newpage
\bibliographystyle{acmref}
\bibliography{references}

\end{document}